\documentclass[epj,twocolumn]{webofc}
\usepackage[varg]{txfonts}   
\usepackage{subfigure}
\usepackage{graphics}
\graphicspath{%
               {./}
}%

\woctitle{ND 2022 International Conference on Nuclear Data for Science and Technology}
\begin{document}

\selectlanguage{english}

\title{Comprehensive investigation of fission yields by using spallation- and (p,2p)-induced fission reactions in inverse kinematics}
%
%

\author{
J. L. Rodr\'{i}guez-S\'{a}nchez\inst{1}\thanks{\emph{Corresponding author: joseluis.rodriguez.sanchez@usc.es}}, 
A.~Gra\~{n}a-Gonz\'{a}lez\inst{1},
J.~Benlliure\inst{1}, 
A.~Chatillon\inst{2}, 
G.~Garc\'{i}a-Jim\'{e}nez\inst{1},
J.~Ta\"{i}eb\inst{2}, 
H.~Alvarez-Pol\inst{1}, 
L.~Atar\inst{3},
L.~Audouin\inst{4}, 
G.~Authelet\inst{5}, 
A.~Besteiro\inst{6}, 
G.~Blanchon\inst{2}, 
K.~Boretzky\inst{7},
P.~Cabanelas\inst{1},
E.~Casarejos\inst{6}, 
J.~Cederkall\inst{8},
D.~Cortina-Gil\inst{1},
A.~Corsi\inst{5},
E.~De~Filippo\inst{9},
M.~Feijoo\inst{1},
D.~Galaviz\inst{10},
I.~Gasparic\inst{11},
R.~Gernh{\"a}user\inst{12},
E.~Haettner\inst{7},
M.~Heil\inst{7},
A.~Heinz\inst{13},
M.~Holl\inst{13},
T.~Jenegger\inst{12},
L.~Ji\inst{3},
H.~T.~Johansson\inst{13},
A.~Keli\'{c}-Heil\inst{7},
O.~A.~Kiselev\inst{7},
P.~Klenze\inst{12},
A.~Knyazev\inst{8},
D.~K{\"o}rper\inst{7},
T.~Kr{\"o}ll\inst{3},
I.~Lihtar\inst{11},
Y.~A.~Litvinov\inst{7},
B.~L{\"o}her\inst{7},
N.~Martorana\inst{9},
P.~Morfouace\inst{2},
D.~M{\"u}cher\inst{14,15},
S.~Murillo~Morales\inst{16},
A.~Obertelli\inst{3},
V.~Panin\inst{7},
J.~Park\inst{17},
S.~Paschalis\inst{16},
A.~Perea\inst{18},
M.~Petri\inst{16},
S.~Pietri\inst{7},
S.~Pirrone\inst{9},
L.~Ponnath\inst{12},
A.~Revel\inst{5},
H.-B.~Rhee\inst{3},
L.~Rose\inst{16},
D.~M.~Rossi\inst{3}, 
P.~Russotto\inst{9},
H.~Simon\inst{7},
A.~Stott\inst{16},
Y.~Sun\inst{3},
C.~S\"{u}rder\inst{3},
R.~Taniuchi\inst{16},
O.~Tengblad\inst{18},
H.~T.~T\"{o}rnqvist\inst{3,7},
M.~Trimarchi\inst{9},
S.~Velardita\inst{3},
J.~Vesic\inst{19},
B.~Voss\inst{7},
H.~Weick\inst{7}
\and 
the R$^{3}$B Collaboration
}

\institute{IGFAE, University of Santiago de Compostela, E-15782 Santiago de Compostela, Spain
\and CEA Bruyeres le Chatel, Chemin du Ru, 91297, Bruy\`{e}res-le-Ch\^{a}tel, France
\and Technische Universit\"{a}t Darmstadt, Fachbereich Physik, Institut f\"{u}r Kernphysik, 64289 Darmstadt, Germany
\and IPN Orsay, 15 rue Georges Clemenceau, 91406, Orsay, France
\and CEA Saclay, IRFU/DPhN, Centre de Saclay, 91191, Gif-sur-Yvette, France
\and CINTECX, Universidad de Vigo, E-36200 Vigo, Spain
\and GSI Helmholtzzentrum f\"{u}r Schwerionenforschung, Planckstra\ss e 1, D-64291 Darmstadt, Germany
\and Lund University, Lund, Sweden
\and INFN Laboratori Nazionali del Sud, Via Santa Sofia 62, 95123, Catania, Italy
\and LIP and Faculty os Sciences, University of Lisbon, 1000-149 Lisbon, Portugal
\and RBI Zagreb, Bijenicka cesta 54, HR10000, Zagreb, Croatia
\and Technische Universit{\"a}t M{\"u}nchen, James-Franck-Str 1, 85748, Garching, Germany
\and Institutionen f{\"o}r Fysik, Chalmers Tekniska H{\"o}gskola, 412 96 G{\"o}teborg, Sweden
\and University of Guelph, 50 Stone Road E, N1G 2W1, Guelph, ON, Canada
\and Institut f\"{u}r Kernphysik der Universit\"{a}t zu K\"{o}ln, Z\"{u}lpicher Strasse 77, D-50937 K\"{o}ln, Germany
\and School of Physics, Engineering and Technology, University of York, YO10 5DD York, UK
\and Institute for Basic Science, Center for Exotic Nuclear Studies, 34126, Daejeon, Republic of Korea
\and Instituto de Estructura de la Materia, CSIC, E-28006 Madrid, Spain
\and Jozef Stefan Institute, Slovenia
}

\abstract{%
In the last decades, measurements of spallation, fragmentation and Coulex induced fission reactions in inverse kinematics have provided valuable data to accurately investigate the fission dynamics and nuclear structure at large deformations of a large variety of stable and non-stable heavy nuclei. To go a step further, we propose now to induce fission by the use of quasi-free (p,2p) scattering reactions in inverse kinematics, which allows us to reconstruct the excitation energy of the compound fissioning system by using the four-momenta of the two outgoing protons. Therefore, this new approach might permit to correlate the excitation energy with the charge and mass distributions of the fission fragments and with the fission probabilities, given for the first time direct access to the simultaneous measurement of the fission yield dependence on temperature and fission barrier heights of exotic heavy nuclei, respectively. The first experiment based on this methodology was realized recently at the GSI/FAIR facility and a detailed description of the experimental setup is given here.
}
\maketitle
\section{Introduction}
\begin{figure*}
\centering
\includegraphics[width=\textwidth,keepaspectratio]{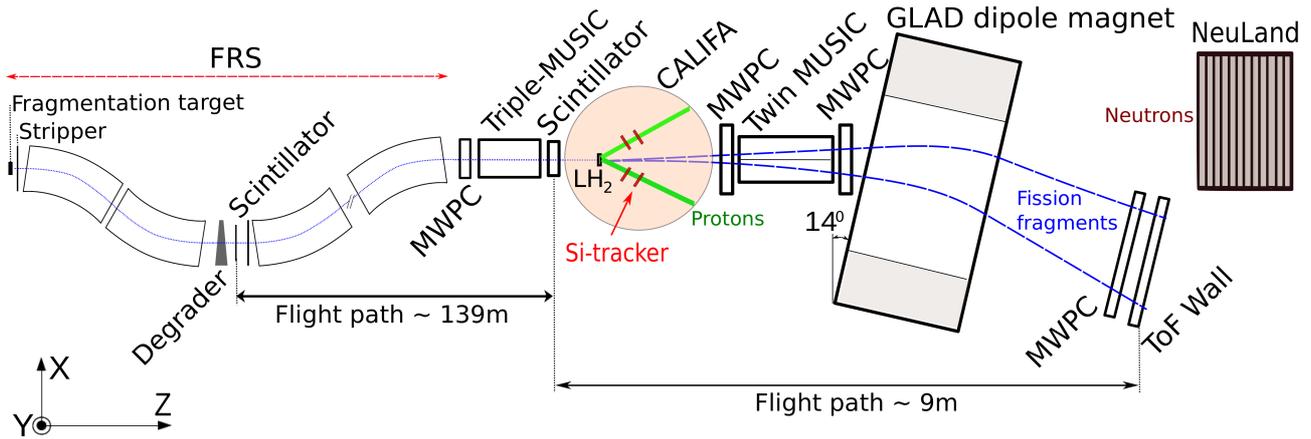}
\caption{(Color online) Top schematic view of the experimental setup used in this work to identify unambiguous the secondary beams and the fission fragments in coincidence with the quasi-free (p,2p) scattering reactions. Sizes are not to scale.
}
\label{fig:setup}
\end{figure*}

Nuclear fission is the clearest example of large-scale collective excitations in nuclei. Since its discovery by Hahn, Meitner, Strassmann and Frisch in 1939~\cite{Meitner1939,Hahn1939}, the progress in the understanding of the fission process has been driven by new experimental results. Despite the recent theoretical progress in the investigation of fission, a complete description still represents a challenge in nuclear physics because it is a very complex dynamical process, whose description involves the coupling between intrinsic and collective degrees of freedom, emission of light particles and $\gamma$-rays, as well as different quantum-mechanical phenomena~\cite{Moller2001}. Therefore, its investigation requires complex experimental setups that allow for complete kinematics measurements of the fission products.

In the late 90's, with the advance of heavy-ion accelerators, a new generation of experimental approaches for fission studies was developed. The use of the inverse kinematics technique permited for the first time an in-flight identification of fission fragments in charge and mass number. The first measurements based on this technique were performed at the GSI facility in Darmstadt (Germany) using the fragment spectrometer FRS~\cite{Geissel92} to detect and identify one of the two fission fragments in charge and mass number~\cite{Armbruster}. The fission reactions were produced mostly on liquid hydrogen and deuterium targets. The data provided relevant information on the fission process dynamics~\cite{Benlliure2006}, discovery of new isotopes and isomeric states~\cite{Jungclaus2007}, as well as production cross sections of more than 1000 nuclear fission residues~\cite{Enqvist2001,Bernas2003,Casarejos06,Loureiro2019}. The FRS spectrometer was also utilized to produce secondary radioactive beams of neutron-deficient actinides and preactinides between the At and U elements~\cite{Schmidt2000} that impinged onto an active target to induce fission through Coulex and fragmentation reactions. The fission fragments were identified in charge by a double ionization chamber. The charge distribution measurements provided relevant information on the transition from symmetric to asymmetric fission~\cite{Schmidt2000} and on the presaddle fission dynamics~\cite{BJ2004,CS07}.

Recently, a great effort was made by the SOFIA collaboration at GSI~\cite{JL2015_1,Pellereau17} to overcome the restrictions of conventional fission experiments and provide complete isotopic measurements of both fission fragments inducing fission through spallation, fragmentation, and Coulex reactions. A state-of-the-art experimental setup based on a large-apperture dipole magnet, multi-sampling ionization chambers, multi-wire proportional counters, and highly segmented plastic scintillator ToF walls allowed for the first time to simultaneously measure and identify both fission fragments in terms of their mass and atomic numbers, as well as to extract correlations between fission observables sensitive to the dynamics of the fission process~\cite{JL2014,JL2015_2,JL2016_1,JL2016_2} and the nuclear structure at the scission point~\cite{Chatillon19,Chatillon20,Martin21,Chatillon22}. Unfortunately, none of these previous measurements permitted the determination of the excitation energy of the fissioning compound nuclei.

To go a step further, we propose now to induce fission of heavy nuclei by using quasi-free (p,2p) scattering reactions in inverse kinematics, which might allow us to reconstruct the excitation energy of the fissioning compound system by measuring the four-momenta of the two outgoing protons~\cite{Noro2022}. In this new approach, fission is induced by the particle-hole excitations left by the removed proton, whose excitation energy ranges from a few to ten’s of MeV. The measurement of the excitation energy in correlation with the atomic and mass number distributions of the two fission fragments represents a powerful tool to study the evolution of fission yields with the temperature for exotic nuclei. Moreover, the excitation energy can be correlated with the fission probability to obtain fission barrier heights.

The realization of this kind of experiment requires a complex experimental setup, providing an unambiguous identification of the secondary beams and a complete kinematics measurement of fission fragments, light particles, and $\gamma$-rays. Such measurement was recently performed at the GSI/FAIR facility by combining the SOFIA experimental setup with a silicon tracker based on AMS-type detectors~\cite{Alcaraz2008}, the calorimeter CALIFA (CALorimeter for In-Flight detection of $\gamma$-rays and high energy charged pArticles)~\cite{Alvarez14}, and the Large Area Neutron Detector (NeuLAND)~\cite{Boretzky21} developed by the R$^{3}$B collaboration. In the following, a detailed description of the experimental setup and some preliminary results are shown.

\section{Experimental setup}
The experiment was carried out at the GSI/FAIR facility where the SIS18-synchrotron was utilized to accelerate heavy ions of $^{238}$U at relativistic energies around 650$A$ MeV. The secondary beams are produced by fragmentation reactions on a 1032 mg/cm$^{2}$ Be target mounted together with a 223 mg/cm$^{2}$ Nb stripper located at the entrance of the fragment separator FRS, as shown in the top-view schematic representation of Fig.~\ref{fig:setup}. The secondary beams are selected by the FRS operated as a momentum-loss achromatic spectrometer~\cite{Khs1987} and then guided to the experimental area $Cave$-$C$ to produce the fission reactions on a liquid hydrogen target (LH$_{2}$). The measurements were divided into two magnetic settings to study (p,2p)-fission reactions of $^{238}$U and exotic nuclei in the region of $^{199-211}$At.

The experimental setup is divided in two parts, one used to characterize the incoming projectile nuclei and another dedicated to measure the fission products. The first part consists of a multi-wire proportional counter (MWPC)~\cite{MWPC}, a triple multi-sampling ionization chamber (triple-MUSIC) and a plastic scintillator detector used to measure the time-of-flight (ToF) of the incoming projectiles and outgoing fission fragments. The first MWPC and the triple-MUSIC detectors provide the beam identification and its position on the target.

The second part consists of three MWPC detectors, a double multi-sampling ionization chamber (Twin MUSIC), a large acceptance superconducting dipole magnet (GLAD)~\cite{Glad} and a ToF Wall. The Twin MUSIC chamber has a central vertical cathode that divides its volume into two active regions (left and right) that are divided into two sections (up and down). Each section is then segmented in 16 anodes that provide 16 independent energy-loss and drift-time measurements. This segmentation allows us to obtain the atomic number of our fission fragments with a resolution better than 0.4 charge units full width at half maximum (FWHM) and the angles on the plane $X$-$Z$ with a resolution better than 1 mrad (FWHM). MWPCs, situated in front and behind the dipole magnet, provide the horizontal (\textit{X}) and vertical (\textit{Y}) positions of the fission fragments. The MWPCs situated in front of the dipole magnet provide the \textit{X} and \textit{Y} positions with a resolution around 200 $\mu$m and 1.5 mm (FWHM), respectively, while the MWPC situated behind the dipole magnet provides those positions with a resolution around 300 $\mu$m and 2 mm (FWHM), respectively. The ToF Wall is made of 28 plastic scintillators that allow to measure the ToF of the fission fragments with respect to the start signal provided by the plastic scintillator located at the entrance of the experimental setup with a resolution around 40 ps (FWHM)~\cite{Ebran}. The GLAD magnet was set to a magnetic field of $\sim$2.7 T and its gap was filled with helium gas at atmospheric pressure to reduce the energy and angular stragglings of the fission fragments. The magnetic rigidity, velocity and atomic number of each fission fragment will be used to obtain the corresponding mass number (\textit{A}) with an average resolution better than $\Delta A/A \sim 0.6 \%$ (FWHM).

Light-charged particles, such as protons, emitted in coincidence with the fission fragments are identified by the CALIFA calorimeter~\cite{Alvarez14}, which is coupled to a silicon tracker (Si-tracker) to reconstruct the trajectory of the two outgoing (p,2p) protons. The Si-tracker is located in front of the LH$_{2}$ target, which was equipped with two double-plane arms consisting of an array of three AMS-type~\cite{Alcaraz2008} 300 $\mu$m thick double-sided silicon-strip detectors. The CALIFA calorimeter is then surrounding the target area to measure the energy of the protons, which consists of 1504 CsI(Tl) crystal scintillators covering a polar angle range between 22 and 90 degrees. The CALIFA energy resolutions range from 1-2$\%$ for stopped protons to 5-15$\%$ for punch through protons.

Finally, prompt neutrons emitted by the fission fragments are measured by the neutron detector NeuLAND~\cite{Boretzky21} located at the end of the experimental setup. This detector consists of 12 double plastic scintillator planes that provide the neutron kinematics and neutron multiplicities with a detection efficiency higher than 50$\%$.

\section{Charge distributions}

\begin{figure}[b!]
\centering
\includegraphics[width=0.48\textwidth,keepaspectratio]{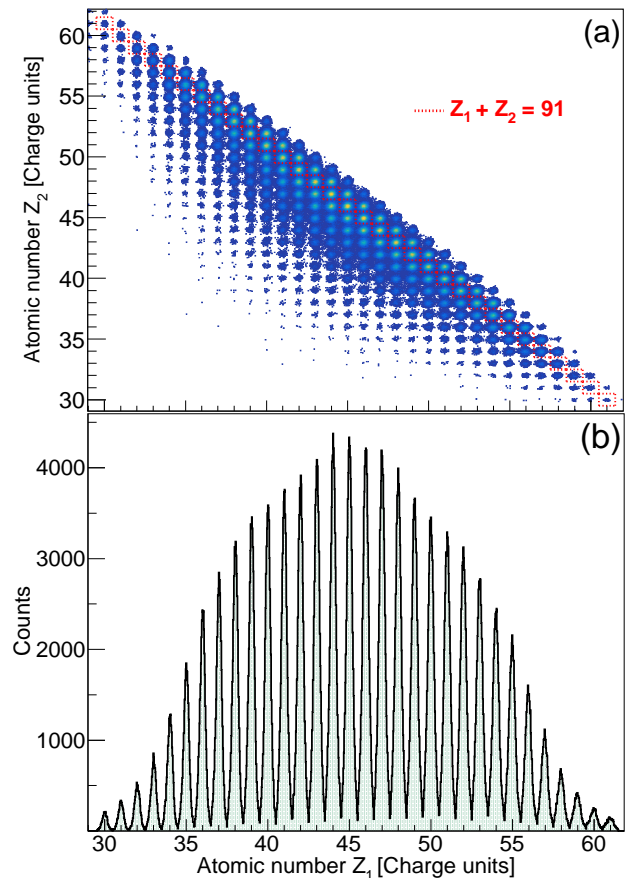}
\caption{(Color online) Charge distribution of fission fragments produced in the reaction $^{238}$U + p at 560$A$ MeV. (a) Correlation plot between the atomic number of the two fission fragments, where the red dotted line indicates the fissioning systems with charge $Z_1 + Z_2 = 91$. (b) Charge distribution of fission fragments for fissioning systems with $Z_1 + Z_2 = 91$.
}
\label{fig:2}
\end{figure}

The atomic number of the fission fragments is deduced based on the fact that the energy loss is proportional to the atomic number squared. In Fig.~\ref{fig:2}(a) we show the measured atomic-number correlation plot in the twin MUSIC detector, where the energy lost by the fission fragments was corrected by the corresponding ToF measurements. The achieved resolution is better than 0.4 charge units (FWHM). The charge distribution for proton removal fission reactions is displayed in the Fig.~\ref{fig:2}(b), which are selected by the condition of fissioning systems with charge $Z_1 + Z_2 = 91$ indicated by a red dotted line in Fig.~\ref{fig:2}(a).

The next step is to employ the CALIFA calorimeter and the Si-tracker for studying the kinematics of the outgoing protons. These detectors will allow to separate the pure quasi-free (p,2p) scattering fission reactions from other reactions involving the simultaneous removal of few neutrons. This measurement will give direct access to study the evolution of fission yields with the excitation energy, which will be explored in future publications. 

\section{Conclusions}
The first (p,2p)-fission experiment has been carried out at GSI/FAIR with projectiles of $^{238}$U and exotic $^{199-211}$At nuclei at relativistic energies of 560$A$ MeV impinging on a liquid hydrogen target. The R$^{3}$B experimental setup is used to perform complete kinematics measurements of all fission products. This measurement will allow us to correlate the fission yields with the excitation energy of the fissioning compound nucleus and to study the energy sharing between the two fission fragments. Additionally, this experiment permits to validate the methodology and develop the data analysis to carry out future experiments with exotic neutron-deficient and neutron-rich heavy nuclei. For the present experiment, the analysis of the charge distribution is finished and the next steps are the identification of fission fragments in mass number and the reconstruction of the excitation
energy spectrum for quasi-free (p,2p) reactions. In the near future further (p,2p)-fission experiments are planned to be carried out at GSI/FAIR with exotic neutron-rich projectiles close to the neutron shell $N=152$ to obtain their fission yields and fission barrier heights for constraining r-process calculations.

\section*{Acknowledgments}
This work was partially supported by the Spanish Ministry for Science and Innovation under grants PGC2018-099746-B-C21 and PGC2018-099746-B-C22, by the Regional Government of Galicia under the program “Grupos de Referencia Competitiva” ED431C-2021-38 and by the “María de Maeztu” Units of Excellence program MDM-2016-0692. J.L.R.S. is thankful for support from Xunta de Galicia under the postdoctoral fellowship ED481D-2021-018.

\end{document}